\documentclass[twoside]{elsart}

\usepackage{amsmath,epsfig,amsfonts,amssymb,latexsym}

\begin{document}
\date{14 March 1998}
\begin{frontmatter}
\title			{Volatility in the Italian Stock Market: an Empirical Study}

\author[INFM,Genova]		{Marco Raberto}, 
\author[INFM,Alessandria]	{Enrico Scalas},
\author[INFM,Dresden]		{Gianaurelio Cuniberti\thanksref{email}}, and
\author[INFM,Genova]            {Massimo Riani}. 
\thanks[email]          {On leave of absence from Dipartimento di Fisica, Universit\`a di Genova, Italy} 
\address[INFM]		{Istituto Nazionale per la Fisica della Materia,
			Unit\`a di Genova}
\address[Dresden]	{Max--Planck--Institut 
			f\"ur Physik komplexer Systeme, \\
			N\"othnitzer Stra{\ss}e 38, 
			D--01187 Dresden, Germany}
\address[Genova]	{Dipartimento di Fisica, 
			Universit\`a di Genova, 
			via Dodecaneso 33, 
			I--16142 Genova, Italy}
\address[Alessandria]	{Dipartimento di Scienze e Tecnologie Avanzate, 
			Universit\`a del Piemonte Orientale, 
			via Cavour 84, 
			I--15100 Alessandria, Italy}

\begin{abstract}
We study the volatility of the MIB30--stock--index high--frequency data from November
28, 1994 through September 15, 1995. Our aim is to empirically characterize the volatility
random walk in the framework of continuous--time finance. To this end, we compute
the index volatility by means of the log--return standard deviation. We choose
an hourly time window in order to investigate intraday properties
of volatility. A periodic component is found for the hourly time window, in agreement
with previous observations. Fluctuations are studied by means of detrended fluctuation
analysis, and we detect long--range correlations. Volatility values are log--stable distributed. 
We discuss the implications of these results for stochastic volatility modelling.
\end{abstract}

\begin{keyword}
Stochastic processes; random walk; statistical finance; econophysics
\\ {{\it PACS: \ }} 02.50.Ey, 02.50.Wp, 89.90.+n
\\ {\it Corresponding author}: Massimo Riani ({\tt riani@ge.infm.it}), 
\\ url: {\tt www.ge.infm.it/econophysics}
\end{keyword}

\end{frontmatter}

\section{Introduction}

In this paper, we study the stochastic properties of MIB30--stock--index volatility.
The Black--Scholes/Merton (BS/M) framework for option pricing is based on the assumption
of constant volatility \cite{Merton 90}. 
In practice, volatility is time dependent and the characterization
of its temporal evolution is one of the main task of the increasing number of physicists
working in the financial field \cite{Stix 98}. This problem has been thoroughly studied
both by economists \cite{Hull 87} and mathematicians \cite{Musiela 97}. 
However, the emphasis has been mainly given to analytically tractable problems. 

Here, we use different approaches in the attempt to empirically characterize
the MIB30--stock--index volatility.
Among the various possible methods, we focus on  the stochastic continuous--time
volatility approach to option pricing.

The paper is divided as follows: in section 2, we present an outline of the 
stochastic--volatility theory; in section 3, the main empirical results are discussed; 
finally, conclusions are drawn in section 4.  

\section{Theory}

In continuous--time finance, stochastic volatility, $\sigma_{t}$, can be modeled by means of two stochastic
differential equations \cite{Musiela 97}, a two factor model:
\begin{equation}
\label{SDE2}
dP(t)=\mu (P,t)P(t)dt+\sigma _{t}P(t)dw_{1}(t)
\end{equation}
\begin{equation}
\label{SDEstoc}
d\sigma _{t}=\alpha (\sigma _{t},t)dt+\beta (\sigma _{t},t)dw_{2}(t)
\end{equation}
where $\mu (P,t)$, $\alpha (\sigma _{t},t)$ and $\beta (\sigma _{t},t)$ are deterministic
functions of the spot price, \( P(t) \), or of the stochastic volatility, $\sigma _{t}$, and of time, \( t \); 
the volatility, \( \sigma _{t} \), is a stochastic variable, \( w_{1} \) and \( w_{2} \)
are standard one--dimensional Brownian motions with correlation 
\( d \langle w_{1}w_{2} \rangle =\rho \, dt \)
for some constant \( \rho  \).
The two processes are independent if and only if \( \rho =0 \) \cite{Musiela 97}.
The problem is then to find a unique solution \( (\tilde{P},\tilde{\sigma} ) \) for the system of stochastic differential
equations (\ref{SDE2}) and (\ref{SDEstoc}).
In the original BS/M model, $\alpha$ and $\beta$ vanish and $\mu$ and $\sigma$ are constant.

If volatility is a stochastic process, continuous riskless hedging in the sense of BS/M (using an option 
and the underlying asset) is not possible \cite{Campbell 97}.
This claim is based on a theorem concerning multi--factor stochastic models.
In the particular case of the two--factor model of eqs. (\ref{SDE2}) and (\ref{SDEstoc}),
in order to form a continuous riskless hedge, a financial instrument with price
fully correlated to volatility would be necessary \cite{Campbell 97}.
A clear and exhaustive introduction to multi--factor models can be found in Marco
Avellaneda's tutorials \cite{Avellaneda web}. 

If the {\em ansatz} of eq. (\ref{SDEstoc}) is accurate,
the empirical analysis of stochastic volatility should lead to the
determination of the coefficients $\alpha(\sigma_{t}, t)$ and $\beta(\sigma_{t}, t)$
\cite{Wilmott 98}, thus completely specifying its stochastic dynamics. 
In practice, this task can be very difficult, due to data incompleteness
and to possible intrinsic mathematical difficulties. For instance, more than one set
of the coefficients could well reproduce the known statistical properties of the volatility
time series.

\section{Empirical Study}

We have analyzed MIB30 high frequency data from 28 November 1994 up to 15 September
1995. MIB30 is an official index of the Italian Stock Exchange, it is composed
by the 30 Italian shares with the highest capitalization and trading volumes, and
is recorded every minute. The data set is composed by over 80,000 data: 420 data
for every trading day. Considering the series of index values,  $P_{j \tau}$, where 
$\tau = 1$ min and  $j=0,\ldots,83579$, we divide our data into non overlapping intervals 
or \emph{time windows} of length $T$.
We choose a \emph{time horizon} or \emph{time scale}, $\Delta t$,  which is an integer 
multiple of $\tau$. We compute the logarithmic returns related 
to every interval $T$ as follows: 
\begin{equation}
\label{logret}
r_{n \Delta t} = \log \frac{P_{(n+1) \Delta t}}{P_{n \Delta t}},\,\,\,n=0,\ldots,N-1;
\end{equation}
where $N$ is such that $T = N \Delta t$.

Financial practitioners define historical volatility as the standard deviation of the 
logarithmic returns \cite{Chriss 97}. Following them, we estimate the volatility for 
every time window as follows:
\begin{equation}
\label{volatility}
\overline{\sigma }=\sqrt{\frac{1}{N-1} \sum_{n=0}^{N-1} [r_{n \Delta t}- \overline{r}]^{2}}
\end{equation}
where $\overline{r}$ is the mean value given by:
\begin{equation}
\label{meanret}
\overline{r}=\frac{1}{N} \sum_{n=0}^{N-1} r_{n \Delta t}
\end{equation}
If $\Delta t$ is measured as a fraction of year, we can define the annualized volatility:
\begin{equation}
\label{anvol}
\overline{\sigma }_{{\rm an}}=\sqrt{\frac{1}{\Delta t}}\; \overline{\sigma }
\end{equation}
In order to investigate intraday properties of volatility, we chose a minutely time horizon and 
an hourly time window. The results of the 1393 annualized volatility estimates are plotted in Fig. 1.
\begin{figure}[ht]
{\centering \resizebox*{1\textwidth}{0.4\textheight}{\includegraphics{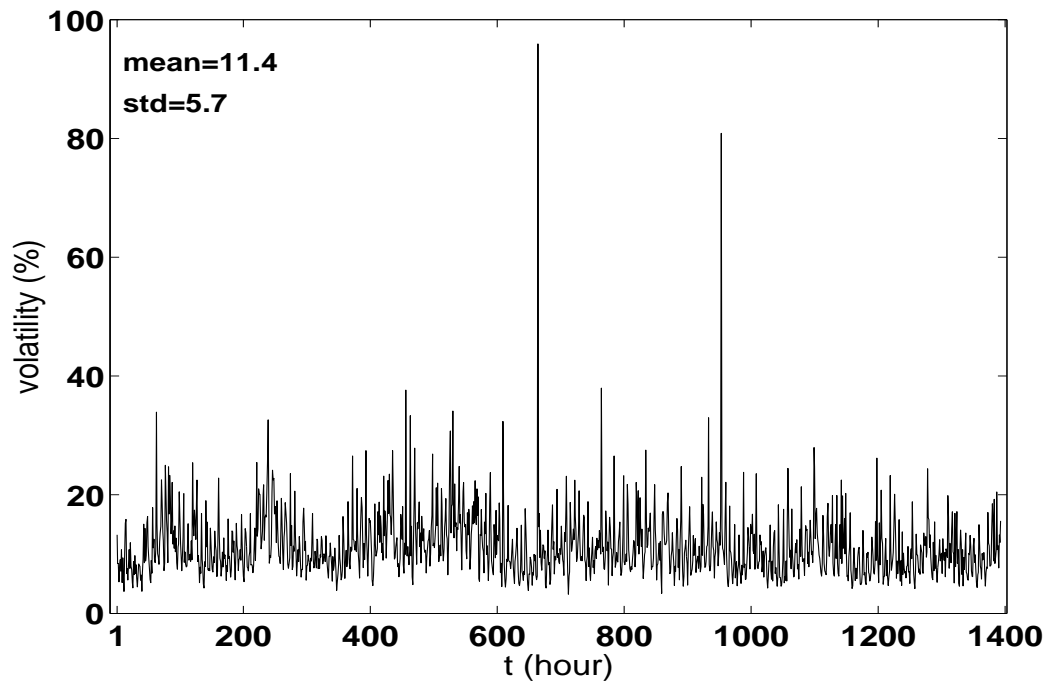}} \par}
\caption{Annualized volatility estimate; the time window is 1 $h$; the time horizon is 1 $min$.}
\end{figure}
 
In Fig. 2 we present the power--spectrum--density estimate computed by means of the correlogram 
method \cite{Marple 87}. 
\begin{figure}[ht]
{\centering \resizebox*{1\textwidth}{0.4\textheight}{\includegraphics{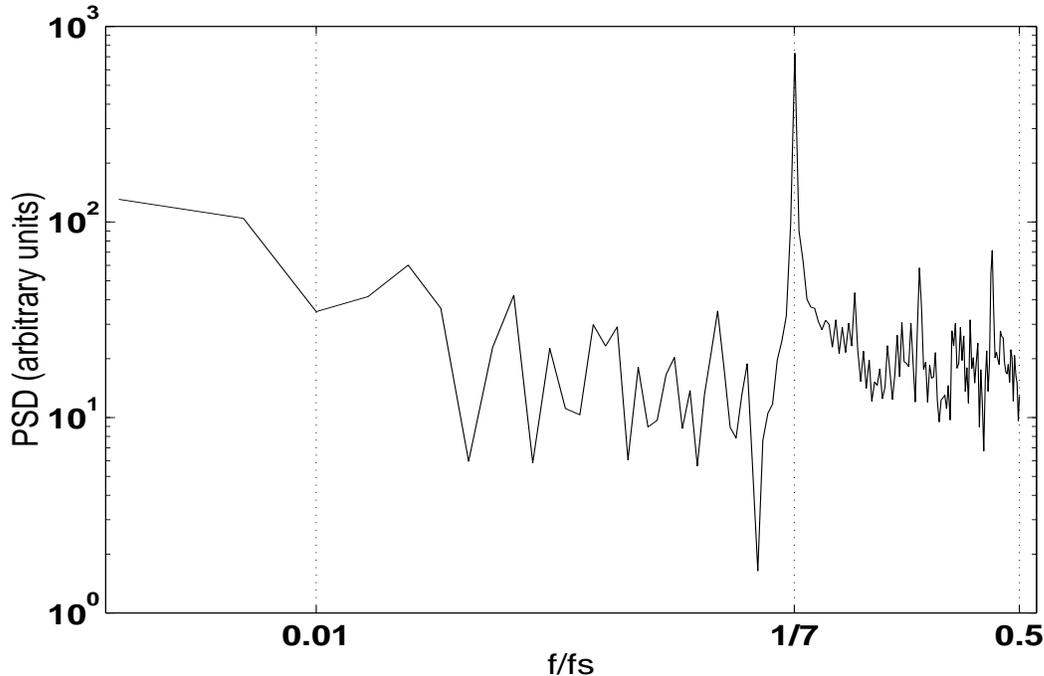}} \par}

\caption{Power spectrum density estimate. The sampling frequency, $f_{s}$, is equal to $1/hour$.}
\end{figure}
The peak
 at $f/f_{s} = 1/7$ is due to a daily periodicity of the volatility values.
In fact, in a day there are seven trading hours and with an hourly time window we get 
seven volatility estimates per day. Indeed, intra--day volatility is U--shaped: it is higher
at the opening and at the closure of the market. This fact has already been observed by 
economists, and it is also known in the physics literature \cite{Cizeau 97}; it probably
reflects the lower trading activity around noon. As a further remark, we are not able to detect 
any low frequency seasonality or observe a clear flicker behaviour at low frequencies \cite{Liu 98}, 
as our time series is less than one year long.

In Fig. 3, an estimate of the volatility probability density function is given. 
We compare the experimental histogram with a log--stable distribution whose parameters, $\delta$ and
$\gamma$ are obtained from empirical data.
\begin{figure}[ht]
{\centering \resizebox*{1\textwidth}{0.4\textheight}{\includegraphics{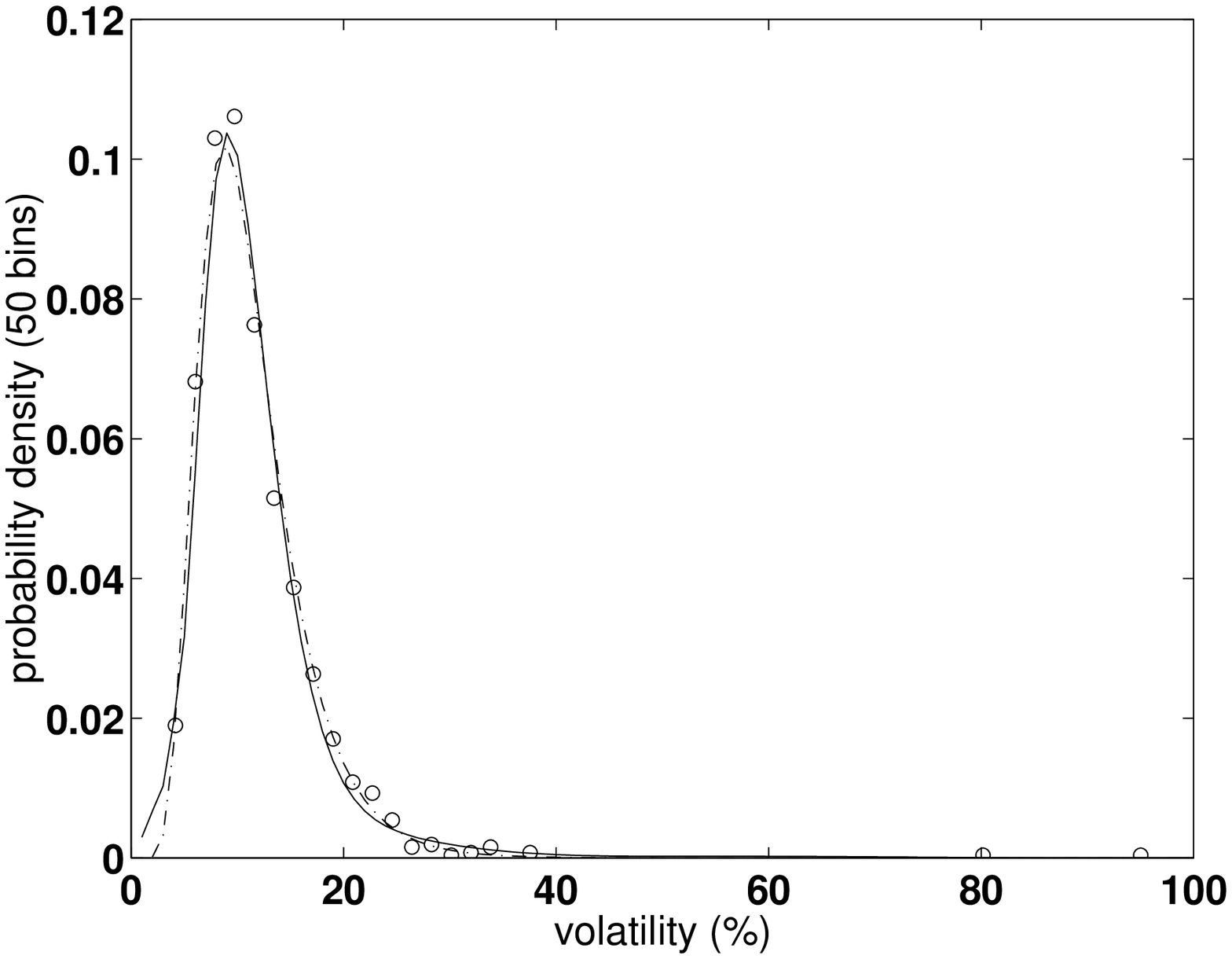}} \par}
\caption{Probability density function estimate; open circles: experimental histogram; solid line: log--L\'evy fit
with $\delta=1.6$ and $\gamma=0.13$; dash--dotted line: log--normal fit with mean value = 2.34 and standard 
deviation = 0.41.}
\end{figure}
Stable or Pareto--L\'evy distributions \cite{Levy 37} have been introduced in the sixties 
in finance and economics \cite{Mandelbrot 63} and their scaling properties have been recently
investigated in relation to the S\&P500 stock index\cite{Mantegna 95}. 
Zero mean stable distributions are described by the following equation:
\begin{equation}  
\label{levy}
P_{\delta, \gamma}(x)=\frac{1}{\pi} \int_{0}^{\infty} \exp (- \gamma q^{\delta}) \cos(q x)\,dq.
\end{equation}
where $\delta=1$ and $\delta=2$ give, respectively, the well known Cauchy and Gauss stable distributions.
A random variable is said to be log--stable distributed if its logarithm follows a stable distribution.

In order to determine the experimental points, the volatility range has been divided into fifty equal 
intervals (bins). From Fig. 1, it can be seen that there are two outliers. If the outliers are
taken into account, a log--L\'evy distribution with exponents $\delta = 1.6$ and $\gamma = 0.13$
gives an acceptable fit of the experimental data ($\chi^{2} \simeq 75$ with 47 degrees of
freedom), whereas the log--normal fit with mean value and standard deviation drawn from data does 
not agree in the tail region ($\chi^{2} \sim 10^{4}$).
Conversely, if the two outliers are rejected as bad data points, and a new histogram is computed
using twenty bins, the log--normal fit is as good as the log--L\'evy fit (for the log--L\'evy fit, 
$\chi^2$ is 29.2, whereas for the log--normal, it is 28.7; this time there are 17 degrees 
of freedom).   

In Fig. 4, the results of detrended fluctuation analysis (DFA) are presented. DFA is used to 
investigate the presence of correlations in time series \cite{Cizeau 97}, \cite{Liu 98}, 
\cite{Scalas 98}, \cite{Vandewalle 97}. 
\begin{figure}[ht]
{\centering \resizebox*{1\textwidth}{0.6\textheight}{\includegraphics{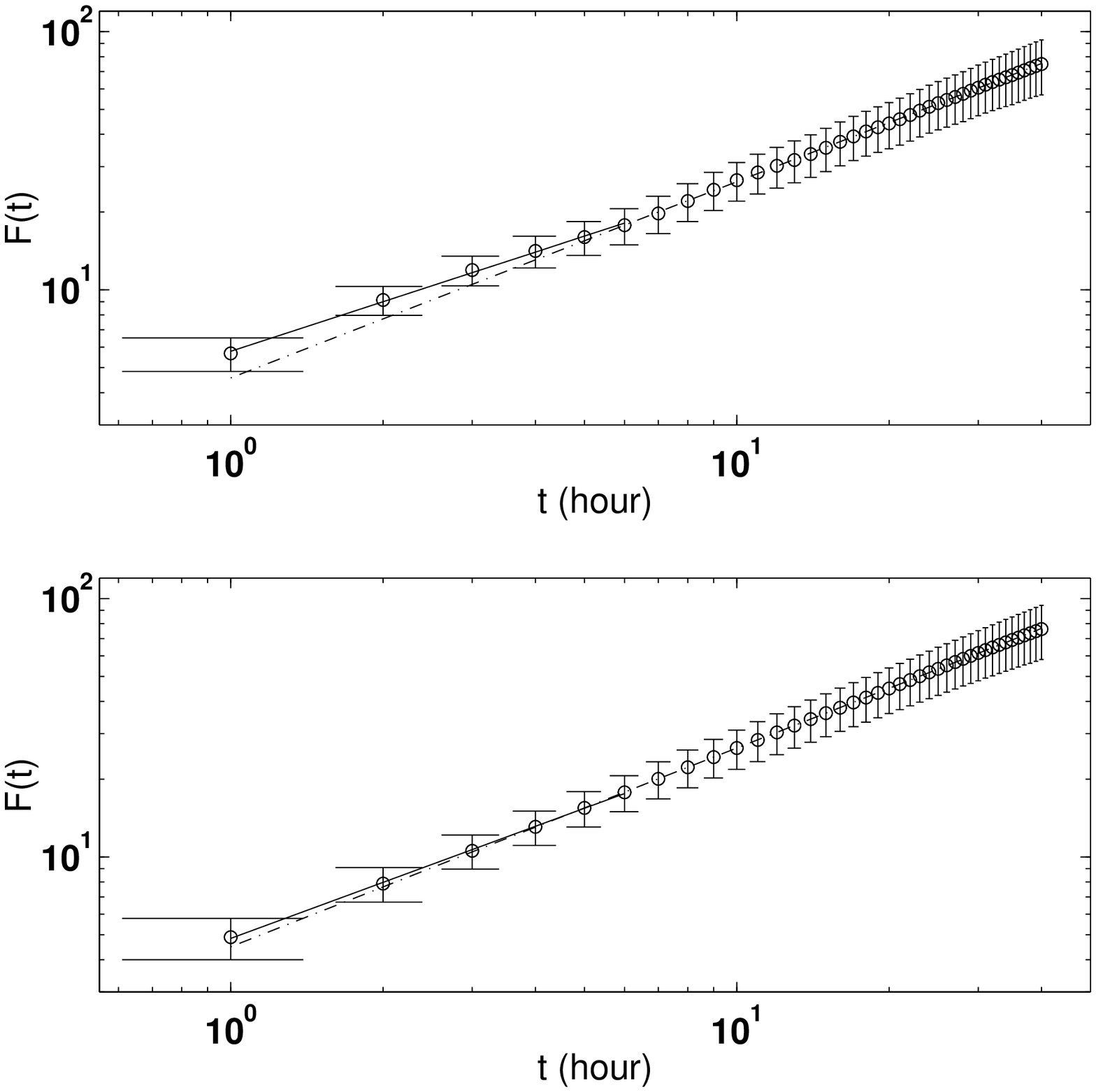}} \par}

\caption{DFA; top: $F(t)$ for the volatility walk; bottom: as top, but for de--U--shaped volatility walk, see 
text for explanation.}

\end{figure}
A volatility walk is defined by means of the displacement $y(t)$:
\begin{equation}
\label{volatilitywalk}
y(t) = \sum_{i=1}^{t} \overline{\sigma} (i).
\end{equation}
The mean square fluctuation, $F(t)$, around the average displacement is given by:
\begin{equation}
\label{DFA}
F(t) = \sqrt{\langle \Delta y(t)^2 \rangle - \langle \Delta y(t) \rangle^2},
\end{equation}
where $\Delta y(t) = y(t_{0}+t) - y(t_{0})$, and $\langle \cdot \rangle$ is the average over all the 
initial steps $t_{0}$.

$F(t)$ follows the scaling law:
\begin{equation}
\label{scaling}
F(t) \propto t^{z}.
\end{equation}
If long--range correlations are present \cite{Montroll 84}, the scaling exponent attains values $z \neq 1/2$.

The points in Fig. 4, top, have been computed according to Eq. (\ref{DFA}) with overlapping subseries.
Forty $F(t)$ estimates are plotted with their error bars. Errors are computed assuming a Gaussian
distribution. The solid line is a least--square linear fit of the first six points, 
corresponding to one trading day; its slope gives $z = 0.64 \pm 0.23$. 
The dash--dotted line is a  least--square fit of the next thirty--four points
and has a slope $z = 0.76 \pm 0.16$. Therefore, from our data we cannot safely conclude that
there is an exponent cross--over at $t = 1$ trading day, as was found in ref. \cite{Liu 98}. However, 
it is possible to argue that $z > 1/2$; indeed, a linear fit of the forty points gives
$z = 0.72 \pm 0.10$.

In Fig. 4, bottom, we present $F(t)$ computed by ``de--U--shaping'' volatility values according to
the following recipe:
\begin{equation}
\label{deushape}
\sigma^{*}(i) = \overline{\sigma}(i) / n(i) \, \, \, \, \, \, \, \, i=1,\ldots,1393;
\end{equation}
where the normalization coefficient, $n(i)$, is given by:
\begin{equation}
\label{normalization1}
n(i) = \frac{1}{q}
\frac{\sum_{k=0}^{q-1} \overline{\sigma} (i+7k)}{\langle \overline{\sigma} \rangle},
\, \, \, i=1,\ldots,7,
\end{equation}
where $q=1393/7$ and $n(i)$ has period 7:
\begin{equation}
\label{normalization2}
n(i+7h)=n(i) \, \, \, \, \, \, \, \, \, \, \, \, h \in \Bbb{Z}.
\end{equation}

In this way, the daily periodicity is removed and the intra--day estimate of $z$ is $0.72 \pm 0.26$, 
whereas
the extra--day estimate is $z = 0.77 \pm 0.16$. The possible cross--over seems to be 
suppressed, and the full linear least--square fit gives $z = 0.76 \pm 0.10$, that is $z > 1/2$.

\section{Summary and Conclusions}

In this paper, we have empirically studied the intra--day statistical properties of stochastic volatility for
the MIB30 index. Stochastic volatility poses serious problems in contingent claim analysis and risk management.
In many cases, risk managers consider a daily time window for computing volatility and perform daily
hedging. However, the preliminary analysis of a recent liquidity crisis of a hedge fund seems to suggest 
that frequent intra--daily hedging could be necessary \cite{The Economist 98}.

Empirical analysis constrains the form of the stochastic differential equations describing 
the time evolution of volatility. 

In particular, we find in the MIB30 index that the
volatility has a periodic behaviour with a one trading day period. Due to the limited 
amount of available data, we are not able to detect any low--frequency seasonality.

Probability--density--function estimates 
indicate that volatility is log--stable distributed; if outliers are taken into account a log--L\'evy
distribution gives a better fit than the usually assumed log--normal distribution \cite{Wilmott 98}.

Finally, the DFA results are compatible with the presence of long--range volatility correlations, but
we cannot safely conclude that there is a crossover between intra--day and extra--day scaling exponents.

\section{Acknowledgements}

The authors are indebted to Rosario N. Mantegna for providing MIB30 data, for suggesting the work 
on stochastic volatility, and for helpful discussions.

\end{document}